\documentstyle[psfig]{aipproc}
\newcommand{\parth}[3]{\left(\frac{#1}{#2\ \rm #3}\right)}

\begin{document}
\title{Mass loss from a magnetically driven wind emitted by a disk
orbiting a stellar mass black hole}
\author{Fr\'ed\'eric Daigne$^{*,**}$ and Robert Mochkovitch$^*$}
\address{$*$Institut d'Astrophysique de Paris,
98 bis boulevard
Arago, 75014 Paris, France\\
$**$Max-Planck-Institut f\"ur Astrophysik,
Karl-Schwarzschild-Str. 1, 85748 Garching, Germany}
\maketitle
\vspace*{-4ex }

\begin{abstract}
The source of cosmic gamma--ray bursts (hereafter GRBs) is usually believed
to be a stellar mass black hole accreting material from a thick
disk. The mechanism for the production of a relativistic wind by such a
system is still unknown. We investigate here one proposal where the
disk energy is extracted by a magnetic field amplified to very large values
$B\sim 10^{15}$ G. Using some very simple assumptions we
compute
the mass loss rate along magnetic field lines and then estimate the
Lorentz
factor $\Gamma$ at infinity. We find that $\Gamma$ can reach high
values only if severe constraints on the field geometry and the
conditions
of energy injection are satisfied. We discuss the results in the
context
of different scenarios for GRBs.
\end{abstract}
\vspace*{-2ex}

\section{Introduction}
Most of the sources which are now discussed to explain
GRBs (the coalescence of two compact objects or the collapse of a
massive star to a black hole (collapsar)
\cite{Narayan92,Woosley93,Paczynski98}) lead to the same
system : a stellar mass black hole surrounded by a thick
debris torus.
The release of energy by such a configuration can come from the
accretion of disk material by the black hole or from the rotational
energy
of the black hole extracted by the Blandford-Znajek
mechanism. 
The released energy is first injected into a
relativistic
wind and then converted into gamma--rays, via the formation of shocks probably
within
the wind itself \cite{Rees94,Daigne98}. The wind is finally
decelerated by the external medium which leads to a
shock responsible for the afterglow emission observed in the
X--rays, optical and radio bands \cite{Wijers97}.

The production of the relativistic wind is a very
complex question because of the very low baryonic load that has to be
achieved in order to reach high values of the terminal Lorentz factor. 
Just a few ideas have been proposed and none appears to be 
fully conclusive. A first possibility to extract the energy
from accretion is the annihilation of neutrino--antineutrino pairs
emitted by the hot disk along the rotation axis of the system, which
is
a region strongly depleted in baryons 
due to 
centrifugal forces. 
The low efficiency of this process however requires high neutrino
luminosities
and therefore short accretion time scales \cite{Ruffert97}. Another
possibility to extract the energy from accretion is to assume that 
the magnetic field in the disk is amplified
by differential rotation to very large values ($B \sim 10^{15}\
\rm G$). A magnetically driven wind could then be emitted from the
disk with a fraction of the Poynting flux being eventually transferred
to matter. The energy can also be extracted from the
rotational energy of the black hole by the Blandford-Znajek mechanism \cite{Lee99}.\\
We present here an exploratory study of the case where a magnetically
driven wind is emitted by the disk. Matter is heated at the basis of the 
wind (by $\nu{\bar \nu}$ annihilation, viscous dissipation, 
magnetic
reconnection, etc.) and then escapes, guided along the magnetic field
lines.
Section \ref{sec_toymodel} describes a ``toy model'' to explore the behavior 
of such a wind. Despite its extreme simplicity, we expect that
it can help to identify the key parameters controlling the baryonic load. Our
results
are presented in section \ref{sec_results} and discussed in 
section \ref{sec_discussion} in the
context of different scenarios for GRBs.
\section{A ``Toy Model''}
\label{sec_toymodel}
We solve the wind equations with the following simplifications : 
(i) we assume a geometrically thin disk and a
poloidal magnetic field with the most simple geometry (straight lines
making an angle $\theta$ with the disk) ; 
(ii) we consider that a stationnary regime has been reached by the wind;
(iii) we use non--relativistic equations (to obtain the mass loss rate we just
need to solve
them up to the sonic point, where $v<0.1 c$) but we
adopt the Paczy\'nski-Wiita potential for the black hole \\
\begin{equation}
\Phi_{\rm BH} = - \frac{G M_{\rm BH}}{r - r_{\rm S}}\ \ \ {\rm with}\ \ \
r_{\rm S} = \frac{2 G M_{\rm BH}}{c^{2}}\ .
\end{equation}

We write the flow equations (continuity, Euler and
energy equations) in a frame corotating with the foot of the
field line, anchored at a radius $r_{0}$ in the disk 
\begin{eqnarray}
\rho\ v\ s(x) & = & \dot{m}\ ,\\
v \frac{dv}{dx} & = & g(x) r_{0} - \frac{1}{\rho} \frac{dP}{dx}\ ,\\
v \frac{d\epsilon}{dx} & = & \dot{q}(x) r_{0} + v \frac{P}{\rho^{2}}
\frac{d\rho}{dx}\ ,
\end{eqnarray}
where $x=\ell / r_{0}$, $\ell$  being the distance along the magnetic field
line, and $\rho$, $P$, $\epsilon$ and $v$ are the density, pressure,
specific internal energy and velocity in the flow.
 
The total
acceleration $g(x)$ includes both gravitational and centrifugal terms.
In this exploratory study the power
deposited per unit mass $\dot{q}(x)$ 
only takes into account the heating and cooling due to neutrinos. 
We assume that the inner part of the disk is optically
thick 
(which is probably justified for 
compact object mergers
but is more
questionnable 
for collapsars
except for low
$\alpha$--viscosity ($\alpha < 0.01$) \cite{Popham99}).
We include the following processes : neutrinos capture on free nucleons, 
neutrino scattering on
relativistic electrons and positrons and neutrino--antineutrino
annihilation (heating); neutrino emission by nucleons and
annihilation of electron--positrons pairs (cooling). 
%
%
The temperature distribution in the disk corresponds to a
gemetrically thin, optically thick disk :
\begin{equation}
T_{\nu}(r) = T_{*} \left(\frac{r_{*}}{r}\right)^{3/4}
\left(\frac{1-\sqrt{\frac{r_{in}}{r}}}{1-\sqrt{\frac{r_{in}}{r_{*}}}}\right)^{1/4}\
{\rm (} T_{*}\ {\rm is\ the\ temperature\ at}\ r_{*}{\rm )}\ ;
\end{equation}
The section of the wind $s(x)$ is easily related to the field geometry
because the field and stream lines are coincident. We adopt the equation
of state computed by \cite{Bethe80} which includes nucleons, 
relativistic electrons and positrons and
photons.\\
%
%
%
 
The acceleration $g(x)$ along a field line is negative up to
$x=x_{1}$ for angles larger than $\theta_{1} \simeq 60^{\circ}$
($60^{\circ}$ is the exact value for a Newtonian instead of
a Paczy\'nski--Wiita black hole potential). For $x > x_{1}$,
$g(x)$ is dominated by the centrifugal force. The sonic
point of the flow is located at a distance $x_{s}$ just below $x_{1}$
(the relative difference never exceeds $1 \%$). 
We solve the flow equations in a classical way by inward integration 
along the field
line. We start at the sonic point by fixing trial values
of the temperature $T_{s}$ and the density $\rho_{s}$ from which we
get the velocity $v_{s}$ and the position $x_{s}$ (from the condition of
regularity at $x=x_{s}$) and then the
value
of the mass loss rate $\dot{m}$. We observe that at some position $x_{cr}$, 
the velocity $v$ begins to fall
off rapidly while $T$ reaches a maximum $T_{max} \le
T_{\nu}(r_{0})$. We adjust $T_{s}$ and $\rho_{s}$ so that $x_{cr}$ is
as close as possible to $0$ and $T_{max}$ to $T_{\nu}(r_{0})$.
\section{Results}
\label{sec_results}
\begin{figure}[!t] 
%
\centerline{\psfig{file=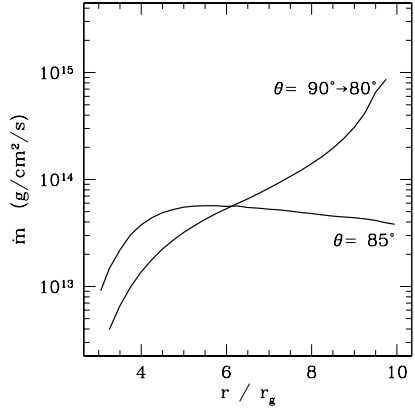,height=5.5cm}}
\vspace*{1ex}

\caption{Mass loss rate from the disk for a constant ($\theta =
85^{\circ}$) and decreasing (from $90^{\circ}$ to $80^{\circ}$ between
$3$ and $10\ r_{\rm g}$) inclination of the field lines. The disk
temperature is $T_{*} = 2\ \rm MeV$ at $r_{*} = 4\ r_{\rm g}$. The mass of
the black hole is $M_{\rm BH} = 2.5\ \rm M_{\odot}$.}
\label{fig_mdot}
\end{figure}

%
%
%
We have studied the dependence of the mass loss rate $\dot{m}$ on the 
different model parameters and found the following expression :
\begin{equation}
\dot{m}(r) \sim 3.8\ 10^{13} \parth{M_{\rm BH}}{2.5}{M_{\odot}}
\parth{T_{\nu}(r)}{2}{MeV}^{10} f\left[\frac{r}{r_{\rm g}};
\theta(r)\right]\ \rm g/cm^{2}/s\ .
\end{equation}
The geometrical function $f$ is normelized in such a way that it is equal
to unity for $r = 4\ r_{\rm g}$ and $\theta(r) = 85^{\circ}$. The very
strong dependance of $\dot{m}$ with $T_{\nu}(r)$ (tenth power) is in
agreement with what is found for neutrino driven winds in spherical
geometry \cite{Duncan86}. Figure \ref{fig_mdot} shows that $\dot{m}$ 
also strongly depends 
on the inclination angle. The other important parameters are
the position in the disk and the mass of the black hole, while
$\dot{m}$ depends only weakly on all other parameters like the size of
the optically thick region (here $r_{\rm in}=3\ r_{\rm g}$ and
$r_{\rm out} = 10\ r_{\rm g}$).
In the more general case where the source of heating is not restricted to
neutrino processes but can also include viscous dissipation, magnetic
reconnection, etc, we have obtained a very simple and general 
analytical approximation for $\dot{m}$ \cite{Daigne2000}
%
%
%
%
\begin{equation}
\dot{m} \sim \frac{\dot{e}}{\Delta \Phi} \delta\ ,
\label{eq_approx}
\end{equation}
where $\dot{e}$ is the rate of energy deposition (in
$\rm erg/cm^{2}/s$) between the plane of the disk ($x=0$) and the sonic
point ($x=x_{s}\simeq x_{1}$), $\Delta \Phi$ is the difference of
potential (gravitational+centrifugal) between $x=0$ and $x=x_{1}$ and  
$\delta$ is a factor close to unity
depending on the distribution of energy injection between $x=0$ and $x=x_{s}$.\\

\vspace*{-1ex} 

We can now estimate the average Lorentz factor 
$\bar{\Gamma} =
\dot{E} / \dot{M} c^{2}$ at infinity. The total
mass loss rate $\dot{M}$ and the power injected into the wind
$\dot{E}$ are given by
\begin{eqnarray}
\dot{M} & = & 2 \int_{r_{in}}^{r_{out}} \dot{m} 2\pi r dr = 2.6\ 10^{26}
\parth{M_{\rm BH}}{2.5}{M_{\odot}}^{3} \parth{T_{*}}{2}{MeV}^{10}\ F_{\rm geo}\
\rm g/s
\label{eq_M}\\
{\rm and}\ \dot{E} & = & 2\ 10^{51} \parth{\Omega/4\pi}{0.1}{}
\parth{f_{\gamma}}{0.05}{}^{-1}
\parth{\dot{E}_{\gamma}}{10^{51}/4\pi}{erg/s/sr}\ \rm erg/s\ ,
\label{eq_E}
\end{eqnarray}
where $F_{\rm geo} = \int_{r_{in}/r_{\rm g}}^{r_{out}/r_{\rm g}} f\left[x;
\theta(x)\right] x dx$ is a function of the field geometry only;  
$\dot{E}_{\gamma}$ is the burst power in gamma--rays,
$\Omega/4\pi$ is the beaming factor and $f_{\gamma}$ is the
efficiency for the conversion of kinetic energy into gamma--rays. 
The wind is powered by accretion but at the same time the disk is
heated by viscous dissipation and cools by emitting neutrinos. 
%
We assume that these losses represent a fraction
$\alpha$ of the power $\dot{E}$ injected into the wind, so that we can
estimate $T_{*}$ at $r_{*} = 4\ r_{\rm g}$ :
\begin{eqnarray}
\dot{E}_{\nu} & = & \alpha \dot{E} = 2 \int_{r_{in}}^{r_{out}}
\frac{7}{8}\sigma T_{\nu}^{4}(r)\ 2\pi r dr\\
{\rm and}\ T_{*} & = & 1.72\ \alpha^{\frac{1}{4}}\ \parth{M_{\rm BH}}{2.5}{M_{odot}}^{-\frac{1}{2}}
\parth{\Omega/4\pi}{0.1}{}^{\frac{1}{4}} \parth{f_{\gamma}}{0.05}{}^{-\frac{1}{4}}
\parth{\dot{E}_{\gamma}}{10^{51}/4\pi}{erg/s/sr}\ \rm MeV\ .
\label{eq_T}
\end{eqnarray}
From equations (\ref{eq_M}), (\ref{eq_E}) and (\ref{eq_T}), we
can calculate the average Lorentz factor 
\begin{equation}
\bar{\Gamma} = \frac{8500}{F_{\rm geo}} \alpha^{-\frac{5}{2}} \parth{M_{\rm
BH}}{2.5}{M_{\odot}}^{2}
\parth{\dot{E}_{\gamma}}{10^{51}/4\pi}{erg/s/sr}^{-\frac{3}{2}}
\parth{\Omega/4\pi}{0.1}{}^{-\frac{3}{2}} \parth{f_{\gamma}}{0.05}{}^{\frac{3}{2}}\ .
\end{equation}
The value of $F_{\rm geo}$ is $56$ for a constant inclination
$\theta=85^{\circ}$ and $250$ if $\theta$ decreases from $90^{\circ}$
to $80^{\circ}$ between $r=3$ and $10\ r_{\rm g}$. 
We therefore conclude that
large terminal Lorentz factors can be reached only if several 
severe 
constraints are satisfied :
(i) low $F_{\rm geo}$ values, i.e. quasi--vertical field lines; (ii)
low $\alpha$ values, i.e. good efficiency for energy injection into
the wind with little dissipation ; (iii) low value of $\Omega / 4\pi$,
i.e. necessity of beaming. 
With the more general equation (\ref{eq_approx}) we can obtain another simple
and useful constraint : if the power $\dot{e}$ deposited below the
sonic point represents a fraction $\chi$ of the total power
$\dot{e}_{\rm tot}$ injected into the wind, we have
\begin{equation}
\Gamma \sim \frac{\dot{e}_{\rm tot}}{\dot{m}c^{2}} \sim
\frac{\Delta \Phi/c^{2}}{\delta \chi}\ .
\end{equation}
For $r=4\ r_{\rm g}$ and $\theta=85^{\circ}$, we obtain $x_{1}=2.182$ and
$\Delta \Phi / c^{2} = 0.18$ which implies that $\chi$ should not
exceed $10^{-3}$ to have $\Gamma > 100$ !
\section{Discussion}
\label{sec_discussion}
This study is clearly limited by its crude assumptions. However the severe 
constraints we get show how difficult it may be to produce
a relativistic MHD wind from the disk. 
An optimistic view of our results would be to consider that this 
difficulty
could just be a way to explain the apparent discrepancy between the observed 
rate of GRBs
and the birthrate of sources in the collapsar scenario, most
collapsars failing to give a GRB. A more pessimistic point
of view would be to conclude that
the baryonic load of such winds is never sufficiently low so that 
they remain non relativistic. 
If one choose to rely on the Blandford-Znajek mechanism to power the wind
\cite{Lee99} it should however be checked that this process is not
"contaminated" by frozen material carried along magnetic field
lines coming from the disk and trapped by the black hole.\\

\end{document}